\documentclass[sigconf]{acmart}
\AtBeginDocument{%
  }

 \acmConference[Conference '26]{Conference}{2026}{}

\setcopyright{none}
\renewcommand\footnotetextcopyrightpermission[1]{}
\usepackage{pifont}
\usepackage{tikz}
\usepackage{algorithm2e}
\usepackage{svg}
\newcommand{\xmark}{\ding{55}}

\usepackage{braket}
\usepackage{subcaption}
\usepackage{amsmath}

\begin{document}

\title{LUNA: LUT-Based Neural Architecture for Fast and Low-Cost Qubit Readout}

\author{Muhammad Ali Farooq}
\orcid{0009-0009-8138-8118}
\affiliation{%
  \institution{Arizona State University}
  \city{Tempe}
  \state{Arizona}
  \country{USA}}
\email{mafarooq19@asu.edu}

\author{Giuseppe Di Guglielmo}
\affiliation{%
  \institution{Fermi National Accelerator Laboratory}
  \city{Batavia}
  \state{Illinois}
  \country{USA}}
\email{gdg@fnal.gov}

\author{Abhi Rajagopala}
\affiliation{%
  \institution{University of Arkansas}
  \city{Fayetteville}
  \state{Arkansas}
  \country{USA}}
\email{abhi@uark.edu}

\author{Nhan Tran}
\affiliation{%
  \institution{Fermi National Accelerator Laboratory}
  \city{Batavia}
  \state{Illinois}
  \country{USA}}
\email{ntran@fnal.gov}

\author{Vidya Chhabria}
\affiliation{%
  \institution{Arizona State University}
  \city{Tempe}
  \state{Arizona}
  \country{USA}
}
\email{vachhabr@asu.edu}

\author{Aman Arora}
\affiliation{%
  \institution{Arizona State University}
  \city{Tempe}
  \state{Arizona}
  \country{USA}
}
\email{aman.kbm@asu.edu}


\begin{abstract}
Qubit readout is a critical operation in quantum computing systems, which maps the analog response of qubits into discrete classical states. Deep neural networks (DNNs) have recently emerged as a promising solution to improve readout accuracy . Prior hardware implementations of DNN-based readout are resource-intensive and suffer from high inference latency, limiting their practical use in low-latency decoding and quantum error correction (QEC) loops. 

This paper proposes \textbf{LUNA}, a fast and efficient superconducting qubit readout accelerator that combines low-cost integrator-based preprocessing with Look-Up Table (LUT) based neural networks for classification. The architecture uses simple integrators for dimensionality reduction with minimal hardware overhead, and employs LogicNets (DNNs synthesized into LUT logic) to drastically reduce resource usage while enabling ultra-low-latency inference. We integrate this with a differential evolution based exploration and optimization framework to identify high-quality design points.

Our results show up to a \textbf{10.95$\times$ reduction in area} and \textbf{30\% lower latency} with \textbf{little to no loss in fidelity} compared to the state-of-the-art. LUNA enables scalable, low-footprint, and high-speed qubit readout, supporting the development of larger and more reliable quantum computing systems.

\end{abstract}



\maketitle

\section{Introduction}
\label{sec:intro}

\begin{figure}[ht]
  \centering
  \includegraphics[width=\linewidth]{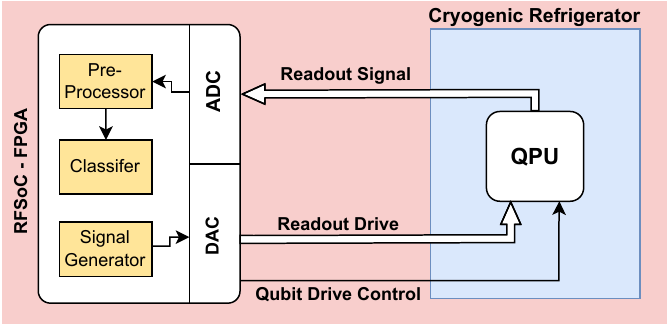}
  \caption{Simplified superconducting-qubit readout chain. RFSoC-FPGA is responsible for qubit drive and readout.}
  \label{fig:readout_chain}
  \vspace{-0.5cm}
\end{figure}

Scalable quantum computing demands fast, accurate, and resource-efficient qubit readout. In superconducting platforms \cite{heinsoo:2018:rapidreadout}, the readout operation converts microwave responses from resonator-coupled qubits into digital I/Q traces, which are then processed and discriminated as $\ket{0}$ or $\ket{1}$. This digital signal processing is typically performed on FPGA or RFSoC-based controllers \cite{yilun:2021:qubic, Keysight2024, ZurichInst2024}, which handle demodulation, integration, and classification in real time (Figure~\ref{fig:readout_chain}).

As quantum processors scale to hundreds or thousands of qubits, readout systems face three constraints: (\textit{i}) \textbf{latency}, since mid-circuit measurement and feedback must occur within coherence windows; (\textit{ii}) \textbf{accuracy}, as assignment errors directly degrade algorithmic and QEC performance; and (\textit{iii}) \textbf{hardware footprint}, as limited FPGA resources must perform many parallel tasks.

Recent work demonstrates that neural-network discriminators can significantly improve readout fidelity by compensating for system nonidealities \cite{lienhard:2022:dnn, maurya:2022:scaling}. However, large models and complex preprocessing stages often make such implementations resource-heavy and slow, limiting scalability and mid-circuit usability. While FPGA-accelerated ML classifiers have shown nanosecond-scale inference \cite{vora:2024:qubicml, guglielmo:2025:fermilab, guo:2025:klinq}, the trade-offs in cost (latency and area) and fidelity remain underexplored.

\begin{table*}[t]
\caption{Representative prior FPGA-based readout efforts. ``DSE/NAS'' indicates whether a formal architecture search was performed.}
\label{tab:related}
\centering
\begin{tabular}{l l l l l l}
\toprule
Work & Qubits & Preprocessing & Classifier & FPGA/RFSoC? & DSE/NAS? \\
\midrule
Lienhard \cite{lienhard:2022:dnn} & 5 & Matched filter & Fully connected DNN & \xmark & \xmark \\
Vora \cite{vora:2024:qubicml} & 1 & Matched filter & Shallow NN & \checkmark & \xmark \\
Di Guglielmo \cite{guglielmo:2025:fermilab} & 1 & — & hls4ml DNN & \checkmark & Partial \\
Guo \cite{guo:2025:klinq} & 1 & Matched filter + averaging & Distilled DNN & \checkmark & Compression only \\
\textbf{This work} & 1 & Integrator (co-designed) & LogicNet (LUT-DNN) & \checkmark & DE-based DSE/NAS \\
\bottomrule
\end{tabular}
\end{table*}

\textbf{This work aims to push qubit-readout acceleration toward the ultra-fast, resource-efficient regime.} We co-design preprocessing and classification for FPGA-based qubit readout and make two key observations. 
First, simple \textit{integrators} can be used in lieu of expensive matched filters \cite{turin:1960:matchedfilter}\cite{guo:2025:klinq} with minimal to no fidelity loss, with low  preprocessing cost and reducing model footprint  through dimensionality reduction.
Second, \textit{LUT-based neural networks} map efficiently to FPGA primitives, providing ultra-low-latency inference with minimal area, for qubit classification. 
We jointly optimize both these components via a combined design-space exploration (DSE) and neural architecture search (NAS) framework using differential evolution.

We implement and evaluate our superconducting qubit readout discrimination accelerator on an AMD/Xilinx FPGA. Our results demonstrate significant reductions in FPGA resource usage and inference latency with no degradation in fidelity, compared to the SOTA implementation \cite{guglielmo:2025:fermilab}.

We make the following contributions in this work:
\begin{enumerate}
  \item We introduce the first ever use of \textbf{LUT-based neural networks (LogicNets \cite{logicnets})} for FPGA-based zero DSP qubit state discrimination acceleration.
  \item We develop a \textbf{lightweight integrator-based preprocessing pipeline} that maintains fidelity at far lower cost resource cost.
  \item We perform a joint \textbf{DSE+NAS} across preprocessing and LogicNet architectures using differential evolution to explore and evaluate trade-offs, marking the first structured NAS attempt using LUT based DNNs.
  \item We present an \textbf{FPGA implementation that is compatible with the Quantum Instrumentation and Control Kit (QICK) \cite{stefanazzi:2021:qick}}, demonstrating upto a $10.95\times$ reduction in area and a $30.9 \%$ reduction in latency at competitive fidelities.
\end{enumerate}

The LUNA flow 
is fully automated, and is available at \textit{blinded}.

\vspace{-0.2cm}
\section{Background and Related Work}
\label{sec:background}

\subsection{Qubit readout for superconducting devices}
Single-shot readout of superconducting qubits converts a quantum state into a classical microwave response. A readout pulse probes a resonator coupled to the qubit; the reflected/transmitted waveform is amplified, down-converted, and digitized to yield time-series in-phase (I) and quadrature (Q) traces. The digital front end, typically a Radio Frequency System on Chip (RFSoc) FPGA,  performs demodulation, preprocessing, and classification to produce a binary assignment ($\ket{0}$ or $\ket{1}$). 


Classical preprocessing choices are matched filtering \cite{turin:1960:matchedfilter}  sliding window averaging. Matched filters maximize signal-to-noise ratio (SNR) under Gaussian noise but require multipliers and memory, which scale poorly when replicated across many channels. Sliding window averaging is hardware-efficient (adders and shifts only) but slightly suboptimal in discrimination performance \cite{guo:2025:klinq}. The front-end design thus exposes a fidelity vs. resource tradeoff critical to scalable FPGA-based readout systems. Keeping this tradeoff in mind, we adopt a novel integrator based strategy for preprocessing.


Thresholding methods for classification degrade under crosstalk, nonstationary noise, and device nonlinearities. Learned discriminators (e.g., Deep Neural Networks (DNN)) improve fidelity, especially across multiplexed channels \cite{lienhard:2022:dnn}. These models can compress and denoise time-series inputs while adapting to experimental non-idealities that fixed templates cannot capture.

Several recent works demonstrate real-time ML-based readout on FPGA/RFSoC hardware. Di~Guglielmo et al. present an end-to-end workflow that couples QICK \cite{stefanazzi:2021:qick} with \texttt{hls4ml} \cite{fahim:2021:hlsml},  achieving low-nanosecond inference but consuming tens of thousands of LUTs and hundreds of DSPs\cite{guglielmo:2025:fermilab}, with no preprocessing or dimensionality reduction. Vora et al. deploy RFSoC-integrated discriminators, utilizing digital local oscillator (DLO) as multiplication strategy for deploying matched filters \cite{vora:2024:qubicml} and integrated it into the QubiC framework \cite{yilun:2021:qubic}.
Guo et al. use distillation to compress large readout networks into smaller FPGA-friendly models \cite{guo:2025:klinq}. They utilize both matched filter and window averaging as pre-processors to form feature vectors for classification. These works validate ML inference on FPGA platforms but tend to rely on multiplier-heavy layers (incurring DSP/BRAM cost) or limited/non-structured design-space search for the preprocessing + classifier co-design.

We use the work of Di Guglielmo et al. \cite{guglielmo:2025:fermilab} as a baseline due to the ready availability of the readout data in the form of \cite{fermilab_data}. Given that qubit response varies from device to device,  we do not draw performance comparisons with other works.

\vspace{-0.4cm}
\subsection{LUT-based networks}

Mapping inference directly into FPGA LUTs eliminates multiplier/ DSP dependence by treating quantized neurons as small Boolean truth tables that can be implemented as native K-input LUT primitives. Early work (LUTNet \cite{wang:2020:lutnet}) showed that trained, binarized operators can be hardened into LUT configurations to yield very area-efficient, low-latency inference engines; subsequent efforts developed toolchains to convert trained networks into LUT masks and to unroll operators for maximal parallelism. LogicNets extends this idea with an explicit hardware–software co-design of a LUT mapped neural network: during training it constrains connectivity (low fan-in), encourages sparsity, and uses quantization so that each neuron’s function can be extracted as one (or a small number of) LUT truth tables, producing a directly deployable, highly-pipelined FPGA netlist \cite{logicnets}. Weightless neural networks (WNNs) and RAM-based classifiers (WiSARD variants and recent LogicWiSARD work) take the table-lookup idea further toward memory-centric, lookup-only inference and can achieve extremely low latency and energy, but they typically trade off accuracy or require different memory/BRAM tradeoffs compared with LUT-mapped quantized networks \cite{susskind:2022:wnn,miranda:2022:logicwisard}.

Practically, LUT-based DNN implementation forces a set of co-design tradeoffs. 
Limiting neuron fan-in (commonly to device LUT sizes, e.g., 6–8 inputs) avoids exponential truth-table growth but often requires deeper or wider DNN topologies or input-partitioning (decomposing a large receptive field across multiple LUTs) to preserve accuracy.
Input-packing and carefully chosen quantization schemes are used to amortize LUT cost.
The standard toolflow is to design a fan-in-aware topology, apply sparsification/pruning and constrained retraining, then extract neuron truth tables for direct LUT instantiation. 
The main benefits of LUT-based DNNs are removal of multiplier/DSP resources, single-cycle (or deeply pipelined) per-layer latency, and highly predictable timing. 

We adopt the LogicNets approach in this work because it explicitly co-optimizes topology, sparsity, and quantization to produce directly mappable LUT truth-tables with a controllable fan-in vs. accuracy tradeoff. Compared with LUTNet \cite{wang:2020:lutnet}, LogicNets \cite{logicnets} provides a clearer training-to-netlist pipeline and tighter topology constraints for predictable area–accuracy tradeoffs, and compared with WNNs \cite{susskind:2022:wnn,miranda:2022:logicwisard}, it preserves closer compatibility with modern quantized DNN accuracy while still delivering the LUT-native, multiplier-free hardware advantages we require.
\vspace{-0.2cm}

\subsection{DSE and NAS}
Automated techniques, like reinforcement learning, Bayesian optimization, and evolutionary algorithms, have been applied to DSE and NAS problems. Differential Evolution (DE) is a simple, gradient-free evolutionary optimizer that adapts readily to mixed encodings and costly black-box evaluations and has been used successfully for NAS-style problems \cite{storn:1997:de, awad:2020:denas}. We adopt DE since it's population-generation and selection mechanism fits naturally with our search; candidate evaluations (DNN training + resource cost prediction) are expensive but can be easily parallelized, as compared to other meta-heuristic approaches like Simulated Annealing.

\begin{figure}[t]
\centering
\includegraphics[width=0.9\linewidth]{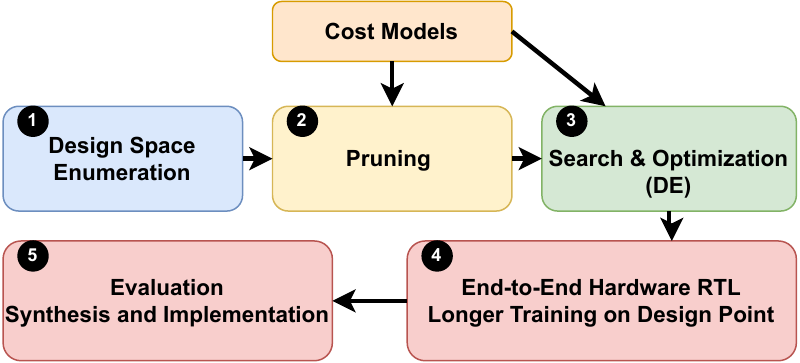}
\caption{LUNA co-design flow: (1) enumerate, (2) prune, (3) DE-based NAS, (4) generate RTL (integrator + LUT-DNN), (5) implement design on target device and evaluate fidelity on test set.}
\vspace{-0.5cm}
\label{fig:overview_flow}
\end{figure}

\subsection{LUNA vs. prior work}
Table~\ref{tab:related} summarizes prior FPGA-based readout works that use ML. Prior efforts demonstrated ML feasibility on FPGA/RFSoC platforms but generally rely on multiplier-based layers, limited preprocessing co-design, or offline compression. To our knowledge, none combine (i) a low-cost integrator front end, (ii) LUT-mapped LogicNet classifiers, and (iii) an automated DSE-NAS that jointly optimizes preprocessing and LUT-DNN topology for area/latency/fidelity tradeoffs.

\begin{figure*}[t]
    \centering
\includegraphics[width=\linewidth]{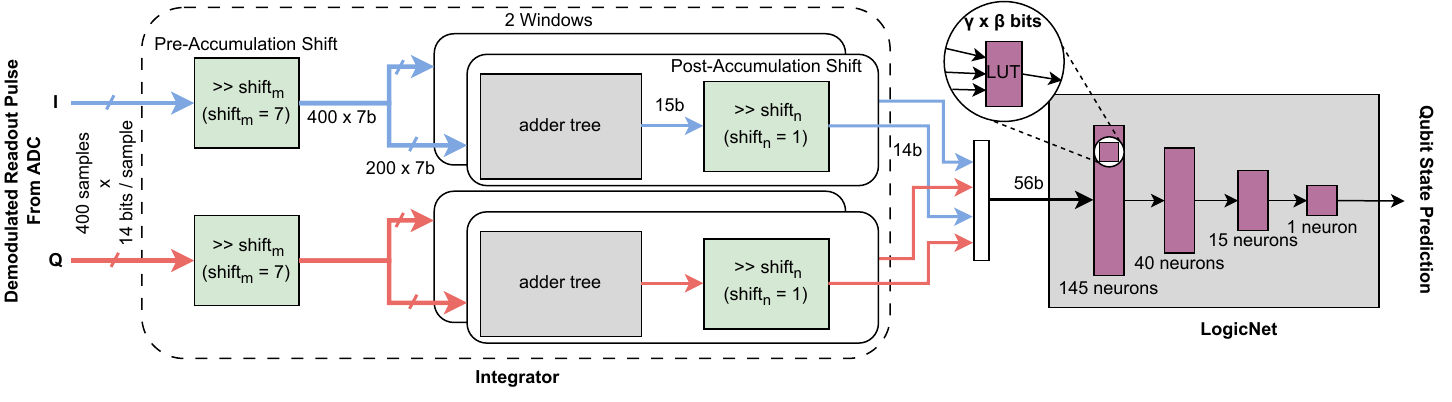}
    \caption{A high level overview of the LUNA architecture. Values shown are demonstrative; taken from our \texttt{fidelity-optimized} solution shown in Table \ref{tab:nasresults}.}
    \label{fig:architecture}
    \vspace{-0.4cm}
\end{figure*}

\section{The LUNA Approach}
\label{sec:proposal}
\subsection{Overview and design flow}
\label{subsec:overview_flow}
Figure~\ref{fig:overview_flow} summarizes our co-design flow. Our approach couples a hardware-aware neural-network design methodology (LogicNets) with an evolutionary architecture search (differential evolution) and automated FPGA synthesis. The flow is as follows: From a parameterized design space \ding{182}, we apply light pruning heuristics to remove infeasible design points \ding{183}, then run a search and optimization loop driven by DE\ding{184}. We then fully implement the resulting design points \ding{185} and implement and evaluate them \ding{186}.

\subsection{Architecture}
\label{sec:arch}
The accelerator comprises two tightly-coupled blocks: (i) an \textbf{integrator preprocessor} that performs low-cost dimensionality reduction, and (ii) a \textbf{LogicNet classifier} where each neuron is mapped directly into FPGA LUTs for ultra-low-latency inference \cite{logicnets}. Data flows in a short, pipelined path: digitized I/Q samples are captured, routed into the integrator windows, reduced to a compact feature vector, then fed into the LogicNet for immediate classification.
A high level overview of this architecture is given Fig. \ref{fig:architecture}.

\subsubsection{Integrator Based Preprocessor}
\label{subsec:integrator}
Integrators compress the raw I/Q trace by partitioning the ADC samples into $\mathsf{num\_filter}$ fixed, non-overlapping windows and computing an accumulation per window. In each window, the incoming 14-bit ADC samples are optionally right-shifted by $\text{shift}_m$ (pre-accumulation) to discard LSB noise and reduce precision, summed in an adder-tree, and then right-shifted by $\text{shift}_n$ (post-accumulation) to normalize dynamic range and produce a quantized scalar I and Q feature per window. These scalar pairs are concatenated to form the input feature vector for the classifier. The integrator is parametrizable by the start ADC sample index/total number of ADC samples, number of windows, and shift amounts; its hardware is an adder-tree with pipeline registers at each stage. We adopt this approach to reduce input dimensionality in order to reduce the cost of the classifier, and to do so in a manner that does not consume expensive DSPs.

\subsubsection{LogicNet classifier}
\label{subsec:logicnet}
The feature vector from the integrator is presented to a LUT-based neural network implemented via LogicNets, where each neuron is synthesized as a Boolean function implemented directly in FPGA LUT primitives \cite{logicnets}. A LogicNet is defined by its layer widths $[\ell_1,\dots,\ell_{k-1}]$, per-layer fanin ($\gamma$) and bitwidth $(\beta)$. Each Neuron Equivalent (NEQ) consumes $X=\beta\cdot\gamma$ input bits and produces $Y=\beta$ output bits, effectively implementing an $X:Y$ LUT operation. These parameters determine both model capacity and FPGA cost. Because all neurons are purely combinational and the network is fully pipelined (typically only 1–2 LUTs between registers) the classifier achieves very low latency suitable for tight QEC timing budgets.

\vspace{-0.25cm}
\subsection{Design space}
\label{sec:dse}
The design space of the integrator and LogicNet is defined by various parameters of each stage, listed in Table \ref{tab:searchspace}.
We model a candidate design point as a fixed-length integer vector $v$ consisting of a value of each parameter.
Each parameter's value can have a large range, leading to a vast design space, which is difficult to explore.
We apply the following heuristics to prune the ranges for each parameter:

\begin{enumerate}
\item \textbf{Area-based pruning:} A conservative LUT estimate is made using a simple model (Section \ref{subsec:area_est}) and designs with estimated LUTs $>A_{\max}=20{,}000$ (empirically chosen) are discarded.
\item \textbf{Full cost-based pruning:} A number of random design points are created after the area-based pruning and fully evaluated  according to the criteria in Section \ref{subsec:cost_eval}. These results are used to further refine our design space.
\end{enumerate}



\vspace{-0.1cm}
\subsection{Search and Optimization}
\label{sec:nas}

The design space of the preprocessing and classifier architecture is very large, due to the large number of variables that govern the architecture (eg. signal window size, normalization, and sparsity settings. See Section \ref{sec:dse}.) The search and optimization of our preprocessing and classifier architecture is unique, as it requires both design space exploration (DSE) for the preprocessor and neural architecture search (NAS) for the classifier. Since isolated optimization of either can ignore better performing points, we are motivated to perform joint optimization of the spaces. We therefore fold the DSE+NAS problem into a single search and optimization problem.

\begin{table}[t]
\centering
\caption{Search-space parameters and allowed ranges.}
\renewcommand{\arraystretch}{1.05}
\setlength{\tabcolsep}{3pt}
\resizebox{\columnwidth}{!}{
\begin{tabular}{lll}
\toprule
\textbf{Parameter} & \textbf{Description} & \textbf{Range} \\
\midrule
\texttt{start time} & Start sample index (ADC); end fixed at 500. & \{0, 50, 100\} \\

\texttt{\# windows} & Signal partitions before preprocessing. & \{1, 2, 3, 4\} \\

\texttt{shift\_m} & Pre-accum. right shift (ignore LSB noise). & \{2, 3, \ldots, 7\} \\

\texttt{shift\_n} & Post-accum. right shift (scale integrator output). & \{0, 1, \ldots, 6\} \\
\midrule
$\ell_0$ & NEQs in input layer. & \{25, 30, \ldots, 145\} \\

\texttt{\# layers} & Hidden layer count. & \{2, 3\} \\

$\ell_1, \ell_2, \ldots$ & NEQs per hidden layer. & \{5, 10, \ldots, 45\} \\

$\boldsymbol{\beta_i, \beta, \beta_o}$ & NEQ input bitwidth (input, hidden, output). & \{1, 2\} \\

$\boldsymbol{\gamma_i}$ & NEQ fan-in (input layer). & \{6, 7\} \\

$\boldsymbol{\gamma, \gamma_o}$ & NEQ fan-in (hidden/output layers). &
$\begin{cases}
\{6,\ldots,16\}, & \beta = 1\\[2pt]
\{6,7,8\}, & \beta = 2
\end{cases}$ \\
\bottomrule
\end{tabular}
}
\label{tab:searchspace}
\vspace{-0.3cm}
\end{table}

We use Differential Evolution (DE) \cite{storn:1997:de} to search the pruned, design space. 
DE  
is a population-based optimization algorithm that we adapt for our mixed-discrete design space. It works by maintaining a population of candidate solutions, which evolve over generations through mutation, crossover, and selection. 
Mutation follows the classical DE rule, where for each member of the population  $v^{(i)}$, we draw three distinct population members $v^{(a)}, v^{(b)}, v^{(c)}$ and form a mutant vector $m$:
\begin{equation}
    m = v^{(a)} + F_{\mathrm{DE}} \cdot \bigl( v^{(b)} - v^{(c)} \bigr)
    \label{eq:de_mutation}
\end{equation}

where $F_{\mathrm{DE}}$ is the mutation rate.
Crossover then constructs a trial `offspring' vector $o$ by combining the mutant $m$ with the parent $v^{(i)}$. Each element in $o$ is determined as:
\begin{equation}
    o_j =
    \begin{cases}
        m_k, & \text{if } k = j, \\
        m_j, & \text{if } u_j < CR, \\
        v^{(i)}_j, & \text{otherwise},
    \end{cases}
    \label{eq:de_crossover}
\end{equation}
where $k$ is a random index selected from the length of the vector, $u_j$ is a random number sampled uniformly in $[0,1]$, and CR is the crossover rate. 
The parameter values of the resulting offspring are clipped and snapped to the nearest valid configuration before evaluation. 
After evaluation, the offspring replaces its parent if it exhibits a lower cost thereby surviving to the next generation. This process repeats for $G_{\max}$ generations, allowing us to arrive at a viable solution point. Early stopping terminates the search when no improvement is observed for a fixed patience window $P$.
With DE, evaluation of each candidate design point becomes  independent and therefore fully parallelizable, making DE an attractive optimization strategy for our use case. With full parallelization across all evaluations within a generation, we are able to evaluate a generation within $\approx4$ minutes.


\subsection{Composite cost and metric estimation}
\label{subsec:cost_eval}

To evaluate each design point, we use a composite cost $\mathcal{C}$ that is defined as a weighted combination of normalized area, latency, and fidelity metrics.
Weights $w_{a}, w_{l}, w_{f}$ define the importance of the three metrics.

\begin{equation}
\mathcal{C}=w_{a}\widetilde{A}+w_{l}\widetilde{L}+w_{f}\widetilde{\mathcal{F}} \text{, where:}
\end{equation}
\begin{equation*}
    \quad
\widetilde{A}=\frac{\text{area}}{A_{\max}}, \quad
\widetilde{L}=\frac{\text{latency}}{L_{\max}}, \quad
\widetilde{\mathcal{F}}=\frac{1-\text{fidelity}}{1-0.90}
\end{equation*}

and where $A_{\max}$ and $L_{\max}$ are empirically chosen (based on the random exploration in Section \ref{sec:de_params}) to be $20{,}000$ LUTs and $14$ cycles. This does not hinder our search process, and allows us to have a meaningful combined cost metric.

The ideal evaluation of a design point would entail training the LogicNet fully on the available data for fidelity evaluation, and complete synthesis and implementation of the end-to-end solution (integrator and LogicNet) on the target device for area and latency. The cost of that design point would then be calculated using the equation above.
However, this would be extremely time consuming, and would make exploration of the large design space prohibitively expensive. As such, we develop inexpensive methods for quick and accurate estimates.

\subsubsection{Latency estimation.}
Latency (cycles) is estimated analytically as
$
\text{latency} = \text{integrator\_cycles} + \text{LogicNet\_stages},
$
with integrator cycles approximated by the adder-tree pipeline depth $\lceil\log_2(N)\rceil$ for $N$ inputs, and each LogicNet layer contributing $1$ pipelined cycle.

\subsubsection{Area estimation.}
\label{subsec:area_est}
Area is approximated by the number of LUTs (flip-flops (FF) are ignored; number of FFs generally tracks with the number of LUTs). LUT count is predicted using a hybrid model:
\begin{itemize}
\item \textbf{Integrator LUTs} are estimated using a regression model trained on a few synthesized adder-tree RTL designs (features: number of inputs, bitwidth).
\item \textbf{LogicNet LUTs} are estimated using analytical per-neuron LUT-equivalents using the analytical formula from \cite{logicnets} and a correction factor obtained through a lightweight regression trained on a few implemented models.
\end{itemize}
This cost model is calibrated using a few end-to-end implementations, and is able to predict end-to-end area within $\pm 20\%$.


\subsubsection{Fidelity evaluation.}
Fidelity $\mathcal{F}$ is obtained by training the LogicNet on the full dataset and reporting the classification metric $\mathcal{F}=1 - 0.5 \cdot \bigl(P(0\mid 1)+P(1\mid 0)\bigr)$ on the test set for parity with prior work for a reduced number of epochs (see Sec.~\ref{sec:training}).

\vspace{-0.2cm}
\subsection{FPGA implementation}
For our final FPGA implementation, we utilize our hand-coded, parameterized RTL templates for the preprocessing segment. For the classifier segment, we use the LogicNets framework \cite{logicnets} to generate RTL from the trained model. These are synthesized and placed-and-routed on the target FPGA. For each design point, we extract post-implementation resource utilization (LUTs, FFs, DSPs) and timing reports (worst negative slack, reported clock period). Similar to \cite{guglielmo:2025:fermilab}, our end to end implementation takes a fixed 2 cycles to save the discrimination prediction to memory; as such, we add a fixed 2 cycles to our final implementation results.

\vspace{-0.2cm}

\section{Methodology}
\label{sec:methodology}

\subsection{Toolchains and platforms}
Our DE implementation and orchestration scripts are implemented in Python~3.8.19. We use the LogicNets framework from AMD/Xilinx \cite{logicnetsgithub} for creating our LUT-based DNN. Parameterized RTL (adder trees and integrators) is generated from Mako templates (mako~1.3.10), enabling a single template family to emit RTL for many design points. Xilinx Vivado~2022.2 performs synthesis and implementation, for timing extraction and to obtain final resource usage.



\vspace{-0.1cm}
\subsection{Dataset}
\label{sec:datasets}
We use the publicly available superconducting-qubit readout dataset (time-series I/Q traces, labeled $\ket{0}/\ket{1}$)  used by Guglielmo et al. \cite{fermilab_data}. The dataset and experimental conventions (readout lengths, sampling, and QICK-based capture) follow the end-to-end readout workflows previously reported.~\cite{guglielmo:2025:fermilab,stefanazzi:2021:qick}

For the experiments reported here, we use a fixed split of 90{,}000 training samples and 10{,}000 test samples. During the search phase (DE evaluation), we train and validate candidate LogicNets on the full split described above to obtain robust fidelity estimates (see next subsection for training protocol). 


\subsection{Training protocol}
\label{sec:training}
To balance search throughput and fidelity estimation reliability, we use a two-stage training protocol:

\begin{enumerate}
  \item \textbf{Search-time training:} During DE search each candidate LogicNet is trained \emph{and} evaluated on the full dataset split for \textbf{5 epochs} using a \textbf{batch size of 512}. This reduced training budget is a trade-off: it yields stable, comparable fidelity estimates across many candidates while keeping per-candidate wall-clock time manageable.
  \item \textbf{Final training:} Once the DE procedure selects a final candidate, that architecture is re-trained from scratch on the same training split for \textbf{30 epochs} using a \textbf{batch size of 1024}. This final training run uses identical preprocessing and quantization settings as during search but with the larger epoch count and batch size to realize the best achievable fidelity prior to FPGA conversion and deployment.
\end{enumerate}

The use of a full-dataset training during search (rather than a proxy subset) helps reduce variance across fidelity estimates when comparing candidate designs.
\vspace{-0.2cm}
\subsection{Differential evolution parameters}
\label{sec:de_params}
We use a population size of $NP=75$ and adopt DE settings of scale factor $F_{\mathrm{DE}}=0.7$ and crossover rate $CR=0.8$. The maximum number of generations is $G_{\max}=150$, and early termination is triggered after $P=40$ generations without improvement. These values were chosen empirically to balance exploration and convergence speed.

\subsection{Baseline}
\label{sec:baseline}
The end-to-end readout work by Guglielmo et al. ~\cite{guglielmo:2025:fermilab} is used as a state-of-the-art reference for ML-based FPGA readout systems, especially since it uses the same dataset \cite{fermilab_data}. As such, we perform the FPGA implementation for the $XCZU49DR$ device as well. 

\begin{table*}[t!]
  \centering
  \caption{Search and Optimization Results. For each optimization target, the top-performing configuration and estimated metrics are listed. Latency is quoted in cycles, Area is quoted in LUTs}
  \label{tab:nasresults}
  \begin{tabular}{l c c c c c c c c c c c c c c}
    \toprule
    & \multicolumn{11}{c}{\textbf{Configuration}} & \multicolumn{3}{c}{\textbf{Estimated Results}} \\
    \cmidrule(lr){2-12} \cmidrule(lr){13-15}
    \textbf{Target} & \textbf{Start Sample} & \textbf{\# Windows} & \textbf{shift$_m$} & \textbf{shift$_n$} & \textbf{Layers} & \textbf{$\beta_i$} & \textbf{$\beta$} & \textbf{$\beta_o$} & \textbf{$\gamma_i$} & \textbf{$\gamma$} & \textbf{$\gamma_o$}  & \textbf{Area} & \textbf{Latency} & \textbf{Fidelity (\%)} \\
    \midrule
    Fidelity & 100 & 2 & 7 & 1 & 145, 40, 15, 1 & 1 & 2 & 2 & 7 & 6 & 8 & 8226 & 12 & 95.995 \\
    Area & 100 & 1 & 9 & 0 & 25, 5, 5, 1 & 1 & 1 & 2 & 6 & 6 & 11 & 6754 & 13 & 95.929 \\
    Latency & 100 & 2 & 9 & 1 & 145, 35, 15, 1 & 1 & 1 & 2 & 6 & 6 & 7 & 7311 & 12 & 95.885 \\
    \bottomrule
  \end{tabular}
  \vspace{-2mm}
\end{table*}

\begin{table*}[t!]
  \centering
  \caption{FPGA Implementation Summary. Results include logic utilization, latency, and fidelity. Resource usage is quoted as percentages of available resources on the $XCZU49DR$ device.}
  \label{tab:fpgasummary}
  \begin{tabular}{llcccccc}
    \toprule
    \textbf{Architecture} & \textbf{Target} & \textbf{LUTs} & \textbf{Flip Flops} & \textbf{DSPs} & \textbf{Period (ns)} & \textbf{Latency (ns)}  & \textbf{Fidelity} \\
    \midrule
    & Fidelity & 10,642 (2.50\%) & 13,206 (1.55\%) & 0 (0.00\%) & 1.751 & 24.51 & 96.059\% \\
    \textbf{LUNA} & Area     & 6,095 (1.43\%)  & 9,688 (1.14\%)  & 0 (0.00\%) & 1.588 &  23.82 & 95.924\% \\
    & Latency  & 6,205 (1.46\%)  & 9,809 (1.15\%)  & 0 (0.00\%) &  1.579 & 22.10 & 95.969\% \\
    \textbf{Di Guglielmo \cite{guglielmo:2025:fermilab}}& --- & 66,600 (15.66\%) & 34,369 (4.04\%) & 0 (0.00\%) & 3.220 & 32.00  & 96.000\% \\
    \bottomrule
  \end{tabular}%
  \vspace{-3mm}
\end{table*}

\section{Results}



\subsection{Search and Optimization Results}
\label{subsec:nasresults}

We perform search and optimization using our described flow under three optimization targets for separate LUNA architectures:
\begin{itemize}
    \item \textbf{Area-Optimized}: prioritize area efficiency \\ ($w_{a}$=0.8,$w_{l}$=0.1,$w_{f}$=0.1)
    \item \textbf{Latency-Optimized}: prioritize inference time \\ ($w_{a}$=0.1,$w_{l}$=0.8,$w_{f}$=0.1)
    \item \textbf{Fidelity-Optimized}: prioritize fidelity \\ ($w_{a}$=0.1,$w_{l}$=0.1,$w_{f}$=0.8)
\end{itemize}

Each configuration reports the best architecture discovered, including integrator and LogicNet parameters, along with estimated performance metrics. The results for each run are given in Table \ref{tab:nasresults}.

The fidelity-optimized model uses a wider network and moderate integrator shifts, reaching  $\approx 96.0\%$  fidelity with a 12-cycle latency and 8226 LUTs. The area-optimized design uses a much smaller LogicNet and more aggressive pre-accumulation, reducing area to 6754 LUTs while retaining 95.93\% fidelity. The latency-targeted run also settles at 12 cycles (7311 LUTs, 95.89\%). The fidelities are tightly clustered while the three points emphasize different area/latency tradeoff --- the search did not produce lower-latency candidates with a lower objective cost, so both latency- and fidelity-targeted runs converged to 12-cycle designs.

\begin{figure}[t]
  \centering
 \includegraphics[width=\linewidth]{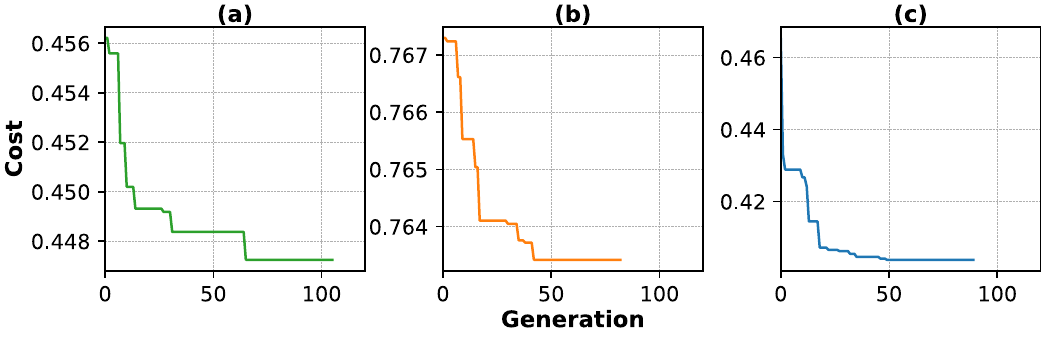}
  \caption{Best cost trajectory across generations for each target objective:
  (a) Fidelity-optimized, (b) Latency-optimized, and (c) Area-optimized.
  Each curve shows the best individual’s cost per generation.}
  \label{fig:cost_traj}
\vspace{-7mm}
\end{figure}

 The cost of the best performing candidate per generation is tracked in Figure \ref{fig:cost_traj}, demonstrating that improvement does occur over the course of DE. 

\vspace{-0.1cm}
\subsection{FPGA Implementation Results}
\label{subsec:fpgaresults}

As summarized in Table~\ref{tab:fpgasummary}, all three target variants achieve large LUT and FF reductions relative to the baseline accelerator while maintaining competitive fidelity and, most notably requiring \textbf{zero DSP blocks}. This demonstrates that both the integrator frontend and the classifier can be mapped entirely onto LUT fabric without sacrificing readout accuracy.

The fidelity optimized LUNA architecture improves upon SOTA fidelity by $0.059 \%$ fidelity, with a $ 6.3\times $ reduction in LUT usage and a $23.4 \%$ reduction in latency. The area optimized architecture achieves a $10.95 \times$ reduction in LUT usage and a $25.6\%$ improvement in latency with only a $0.076\%$ loss of fidelity. The latency optimized architecture improves upon latency by $30.9\%$ with a $10.74 \times$ reduction in LUT use and a fidelity loss of only $0.031\%$. 

These hardware results validate the trends observed during DE-based search and confirm that the flow yields consistently compact and efficient implementations.

\section{Discussion}

This work shows that combining an integrator-based dimensionality reduction stage with LUT-mapped LogicNet classifiers enables substantial FPGA savings while preserving state-of-the-art qubit-state discrimination fidelity. The integrator compresses raw I/Q trajectories into a small set of aggregate features using simple shift-and-accumulate operations, reducing classifier input dimensionality with negligible compute cost. When paired with LogicNets, the resulting LUNA flow achieves order-of-magnitude LUT reductions, lower latency, and zero DSP usage, freeing arithmetic resources for qubit control and signal-generation tasks.

Reducing per-qubit area is increasingly important as quantum systems move toward mid-circuit measurement and quantum error correction (QEC), where large numbers of concurrent readout paths are required. A smaller hardware footprint directly increases the number of readout engines that can be instantiated on a single device, supporting the scaling needs of future processors.

Several avenues for future work follow. 
Evaluating LUNA on additional datasets, including multi-qubit or higher-dimensional readout traces, will help assess generality. Extending the architecture to multi-qubit joint classifiers could capture crosstalk and correlated noise, a limitation of per-qubit pipelines noted in prior work. Finally, richer search strategies may provide improved coverage of the design space and may uncover even better solutions.

\vspace{-0.1cm}
\section{Conclusion}
This paper presents LUNA, a hardware–software co-design framework that couples integrator-based dimensionality reduction with LUT-based neural network classifiers to produce highly efficient FPGA implementations for qubit-state readout. Using differential evolution to jointly optimize preprocessing and classifier structure, LUNA identifies compact designs that maintain high discrimination fidelity while drastically reducing hardware footprint. Across the studied dataset, LUNA reaches fidelities near 96\% with up to a $10.95\times$ LUT reduction and up to a 30\% latency improvement, all with zero DSP utilization. These resource savings directly support scalable quantum processors, particularly scenarios requiring large numbers of parallel mid-circuit measurements and QEC operations.

\bibliographystyle{ACM-Reference-Format}
\bibliography{software.bib}
\end{document}